\documentclass[twocolumn]{aastex63}

\newcommand{\simlt}{{}_{\sim}^{<}\,}
\newcommand{\simgt}{{}_{\sim}^{>}\,}

\received{}
\revised{}
\accepted{}
\submitjournal{ApJL}

\shorttitle{Magnetic First Stars}
\shortauthors{Koh et al.}
\graphicspath{{./}{figures/}}

\begin{document}

\title{First Star Formation in the Presence of Primordial Magnetic Fields}

\correspondingauthor{Tom Abel}
\email{tabel@stanford.edu}

\author{Daegene Koh}
\affiliation{Kavli Institue for Particle Astrophysics and Cosmology\\ Stanford University, SLAC National Accelerator Laboratory\\ Menlo Park, CA 94025}

\author{Tom Abel}
\affiliation{Kavli Institue for Particle Astrophysics and Cosmology\\ Stanford University, SLAC National Accelerator Laboratory\\ Menlo Park, CA 94025}

\author{Karsten Jedamzik}
\affiliation{Laboratoire d'Univers et Particules de Montpellier, UMR5299-CNRS,
Universit\'e de Montpellier, 34095 Montpellier, France}

\begin{abstract}
It has been recently claimed that primordial magnetic fields could relieve the
cosmological Hubble tension. We consider the impact of such fields on the formation of the first cosmological objects, mini-halos forming stars, for present-day field strengths in the range of $2\times 10^{-12}$ - $2\times 10^{-10}$ G. These values correspond to
initial ratios of Alv\'en velocity to the speed of sound of
$v_a/c_s\approx 0.03 - 3$. We find that when $v_a/c_s\ll 1$, the effects are modest. However, when $v_a\sim c_s$, the starting time of the gravitational collapse is delayed and the duration extended as much as by $\Delta$z = 2.5 in redshift. When $v_a > c_s$, the collapse is completely suppressed and the mini-halos continue to grow and are unlikely to collapse until reaching the atomic cooling limit. Employing current observational limits on primordial
magnetic fields we conclude that inflationary produced
primordial magnetic fields could
have a significant impact on first star formation, whereas post-inflationary
produced fields do not.


\end{abstract}

\keywords{stars: Population III --- 
(cosmology:) dark ages, reionization, first stars --- stars: magnetic field --- magnetohydrodynamics (MHD)}


\section{Introduction} \label{sec:intro}

Magnetic fields are observed throughout the local Universe, in galaxies,
in clusters of galaxies, as well as very possibly in the extra-galactic 
medium. Observations of TeV blazars \citep{Neronov:1900zz} 
are most easily explained
by the existence of an almost volume filling magnetic field permeating
the space between galaxies. 
Recently it has also been shown by \citet{jedamzik2020relieving} that magnetic fields of $\sim 0.05\,$nG
existant before the epoch of recombination show good promise to alleviate 
the $4-5\sigma$ cosmic Hubble tension and the $\sim 2\sigma$ cosmic $S_8$
tension within standard $\Lambda$CDM. The Hubble tension is the mismatch
between the inferred small present day Hubble constant $H_0$ 
from observations of the cosmic microwave background radiation by the Planck satellite when assuming $\Lambda$CDM \citep{2018arXiv180706209P},
and a larger $H_0$ inferred by local observations 
\citep{Reid:2019tiq,Wong:2019kwg,Pesce:2020xfe}.
The $S_8$ tension is the difference between the 
$\Lambda$CDM\
predicted current matter fluctuation amplitude on a scale of 8 Mpc $h^{-1}$ and
that observed directly via weak lensing 
\citep{Abbott:2017wau,Asgari:2020wuj}. It has been shown that
weak magnetic fields induce density fluctuations on small, sub-Jeans 
comoving $\sim$ kpc scales before recombination \citep{Jedamzik:2013gua} which would indeed alter the prediction for $H_0$ and $S_8$
within $\Lambda$CDM in a favorable way. Such fields necessarily would be of 
primordial origin.


There are two conceptually different possibilities for the generation of primordial magnetic fields (PMFs, hereafter),
magnetogenesis during inflation and 
post-inflationary generation (often referred to as "causal" scenarios) as for example during a
first-order electroweak phase transition. Though multiple proposals 
exist (cf. to 
\citet{2013A&ARv..21...62D,Subramanian:2015lua,Vachaspati:2020blt} for reviews),
there is no preferred candidate. Inflationary scenarios have to lead to an
approximate scale-invariant magnetic spectrum to be successful, whereas most causal scenarios develop a very blue Batchelor spectrum, with most magnetic power on small scales \citep{Durrer:2003ja,Saveliev:2012ea}. Stringent upper limits on PMFs from observations of the cosmic microwave background radiation have been placed by \citet{2019PhRvL.123b1301J} 
at $\simlt 10^{-10}$G and 
$\simlt 2\times 10^{-11}$ for inflationary and causal fields, respectively.



In this {\em Letter}, we entertain the idea that a primordial magnetic field indeed existed and make steps to investigate their impact on first structure formation.
In $\Lambda$CDM structure is built bottom up, marginal differences in the formation of the first objects could have significant impact for all subsequent structure formation. 
Population III stars were formed at z $>$ 20 in mini-halos of mass $\sim 10^{5-6} M_\odot$ which gravitational collapse via $H_2$ cooling \citep{1997ApJ...474....1T,2002Sci...295...93A}. Magnetic fields are understood to impact present day star formation in a number of ways \citep{2007ARA&A..45..565M}. Although the exact nature of their impact on Pop III stars is still yet unknown, there have been a number of 
somewhat idealized studies exploring these effects \citep{2020MNRAS.496.5528M}, including reducing fragmentation \citep{2020MNRAS.tmp.2043S}, increasing ionization degree \citep{2019MNRAS.488.1846N}, and angular momentum transport \citep{2013MNRAS.435.3283M}. Any weak initial field will be amplified by the small scale dynamo \citep{Xu2008,Sur:2010,Schleicher:2011,Turk:2012aa}, but tends to not strongly alter the initial collapse forming primordial stars. 

On the other hand, \citet{2020arXiv200505401S} considered the impact of PMFs on the total matter spectrum and the subsequently formed dwarf galaxies. They ruled out the highest strengths, (i.e. $0.2-0.5\,$nG for inflationary fields) for a number of reasons. First, the dwarf galaxies in these cases overproduce stars, in contrast to local scaling relations. They also produce enough ionizing photons to reionize the universe prior to z=9, also contradicting numerous other measurements. The properties of dwarf galaxies depend on the chemo-dynamical environment that they are embedded within. As a result, these results cannot be fully interpreted without explicitly resolving the mini-halo scales where the first objects were formed.

The latter is what we attempt here. In addition, we evolve the matter perturbations and magnetic fields from the linear regime at high redshift considering the earliest collapsing object in a fairly large volume. This is in contrast to \citet{2013MNRAS.435.3283M,2019MNRAS.488.1846N,2020MNRAS.496.5528M,2020MNRAS.tmp.2043S} which adopt non-linear initial conditions at lower redshift. However, we stress that unlike \citet{2020arXiv200505401S} we do not take into account the additional power in the baryon density fluctuations generated by the magnetic fields themselves \citep{Wasserman,Kim:1994zh,Subramanian:1998fn}. As much as the above mentioned works, our study can therefore not reach ultimate conclusions but should add an important element to the discussion. 
In the following section, we introduce our simulation setup. Then, in section \ref{sec:results}, we describe the results. Finally, we 
discuss potential impact and caveats in section \ref{sec:con}.

\section{Simulation Setup} \label{sec:simsetup}

Our exploration involves a set of 5 cosmological simulations using the latest public version of the adaptive mesh refinement ideal MHD simulation code Enzo v2.6 \citep{2014ApJS..211...19B,2019JOSS....4.1636B}. Our basic setup is taken from \citet{2016MNRAS.462...81K} with a nested initial grid focusing on a single mini-halo in a 250 $h^{-1}$ comoving kpc box. We consider the same physics as with this earlier study including a nine-species non-equilibrium chemistry model \citep{1997NewA....2..181A}, a time-dependent Lyman-Werner background \citep{2012ApJ...745...50W}, and a radiative self-shielding model \citep{2011MNRAS.418..838W}. Our main focus is the evolution of the mini-halo prior to self-collapse and thus, terminate the simulation when a maximal refinement level of 15 (a spatial resolution of $\sim 400$ astronomical units) is reached and do not follow the subsequent star formation and feedback processes.

In contrast to this earlier study, we modify the following parameters. First, the Jeans length criterion is reduced down to 32
zones per Jeans length. In studies of magnetic field amplification by gravitational turbulence the required minimum for this parameter is 30 to capture the small scale dynamo \citep{2011ApJ...731...62F,Turk:2012aa}. 

The central variable of interest is the initial magnetic field strength. We seed the entire simulation domain with a magnetic field initially pointed in the z-direction. To contextualize the values of the magnetic field strengths chosen, we refer to the speed ratio $v_a/c_s$, where $v_a$ is the Alfv\'en speed, and $c_s$ is the sound speed, and choose the following ratios for our study

\begin{itemize}
    \item 0.03 , 0.30, 0.66, 1.00,  3.00
\end{itemize}

For brevity, we will, in the rest of the {\em Letter}, refer to the ratios $<$1, $\sim$1, $>$1 as sub-sonic, trans-sonic, and super-sonic ratios respectively.
\begin{figure*}[t!]
	\includegraphics[width=\textwidth]{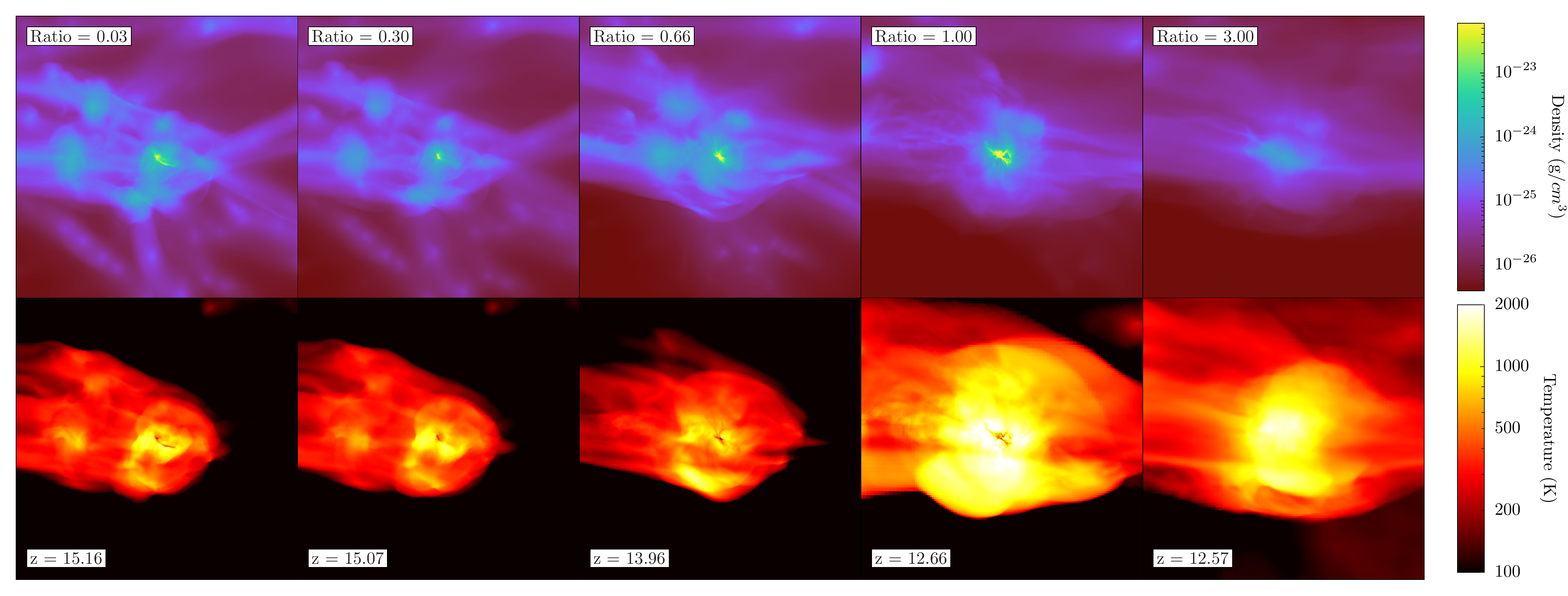}
	\caption{Density-weighted projection plots of density (top) and temperature (bottom) of the different runs all spanning 1 kpc wide. From left to right, the plots are of increasing Alfv\'en to sound speed ratios corresponding to 0.03, 0.30, 0.66, 1.00, and 3.00 respectively. Each plot is taken from the final data output produced at the time of collapse except for the right-most panel which was terminated at a comparable time with the trans-sonic (second from the right) case at z=12.6. }
	\label{fig:multi}
\end{figure*}
A speed ratio of 1 then is equivalent to an initial proper B-field strength, B = $v_a * \sqrt{4 \pi \rho}$ = 1.32 x $10^{-6}$ G at z = 150, where $\rho$ is the mean density in the box. This field strength corresponds to a comoving field strength of  5.79 x $10^{-11}$ G. So the full range of initial comoving field strengths spans 1.75 x $10^{-12}$ G to 1.73 x $10^{-10}$ G. The comoving region from which the $\sim 10^6 M_\odot$ halo forms is approximately 10 kpc across and hence is the relevant scale over which we assume the magnetic field to be initially uniform. 

\section{Results} \label{sec:results}
We analyze the numerical simulations based on full snapshots stored on disk for every 10 Myrs of the evolution until the highest level of refinement of 15 is reached for the first time at which the simulation is terminated and a final data output is produced. The exception is the super-sonic ratio of 3.0 in which the strong B-field inhibited the collapse and we terminated the simulation at z = 12.6.

\subsection{Central Halo Inspection}
Here we show plots comparing the different runs. Figure \ref{fig:multi} shows projection plots of density and temperature of the various runs, each panel spanning 1 kpc across. From left to right, the plots are in order of increasing $v_a/c_s$ ratios, shown at the time when the highest refinement level is reached, corresponding to collapse of the central region. The 0.03 and 0.30 sub-sonic ratio central halos collapse at z = 15, the 0.66 halo at z =14, and the trans-sonic halo at z =12.7. In comparing the two sub-sonic cases of ratios 0.03 and 0.30 where the collapse times are similar, we can see that the extended filament protruding from the central object is less dense in the greater ratio case. Also the surrounding satellite gas clouds all have noticeably reduced densities.

As the time of collapse is delayed at higher $v_a/c_s$ ratios, the central region has more time to grow both in size as it merges with the nearby sub-halos and in temperature. In particular, for the trans-sonic case where $v_a \sim  c_s$, we see an extended temperature cavity where the central halo has collapsed, surrounded by highly heated gas, with temperatures reaching several $T \sim 10^4$ K. The nearby gas clouds to the left and below the central halo in the plane of the plot in the leftmost panel have completely merged with the central halo in the trans-sonic case resulting in significantly elevated temperatures. By the time these clouds merge in, the central halo had already been cooled to form a dense core. The infalling gas then collides with the dense core and is scattered around it producing the extended sub-structure shown in the plot. In the super-sonic case, we see continued heating of the central region of the halo while the density remains quite low and that cooling has not begun even at this late stage. 

One significant trend is that the time of collapse is delayed as a function of increasing speed ratio. The additional magnetic pressure heats the gas and adds an additional barrier for the gravity to overcome to initiate self-collapse. This trend is such that already the trans-sonic halo collapses at z = 12.66 where the magnetic field is contributing to a delay time of $\Delta$z = 2.5. Following this trend, we estimate that the central halo in the highest ratio run of 3.0 is unlikely to collapse even until z = 10 and may only collapse once reaching the atomic cooling limit.   

\begin{figure}
	\includegraphics[width=\columnwidth]{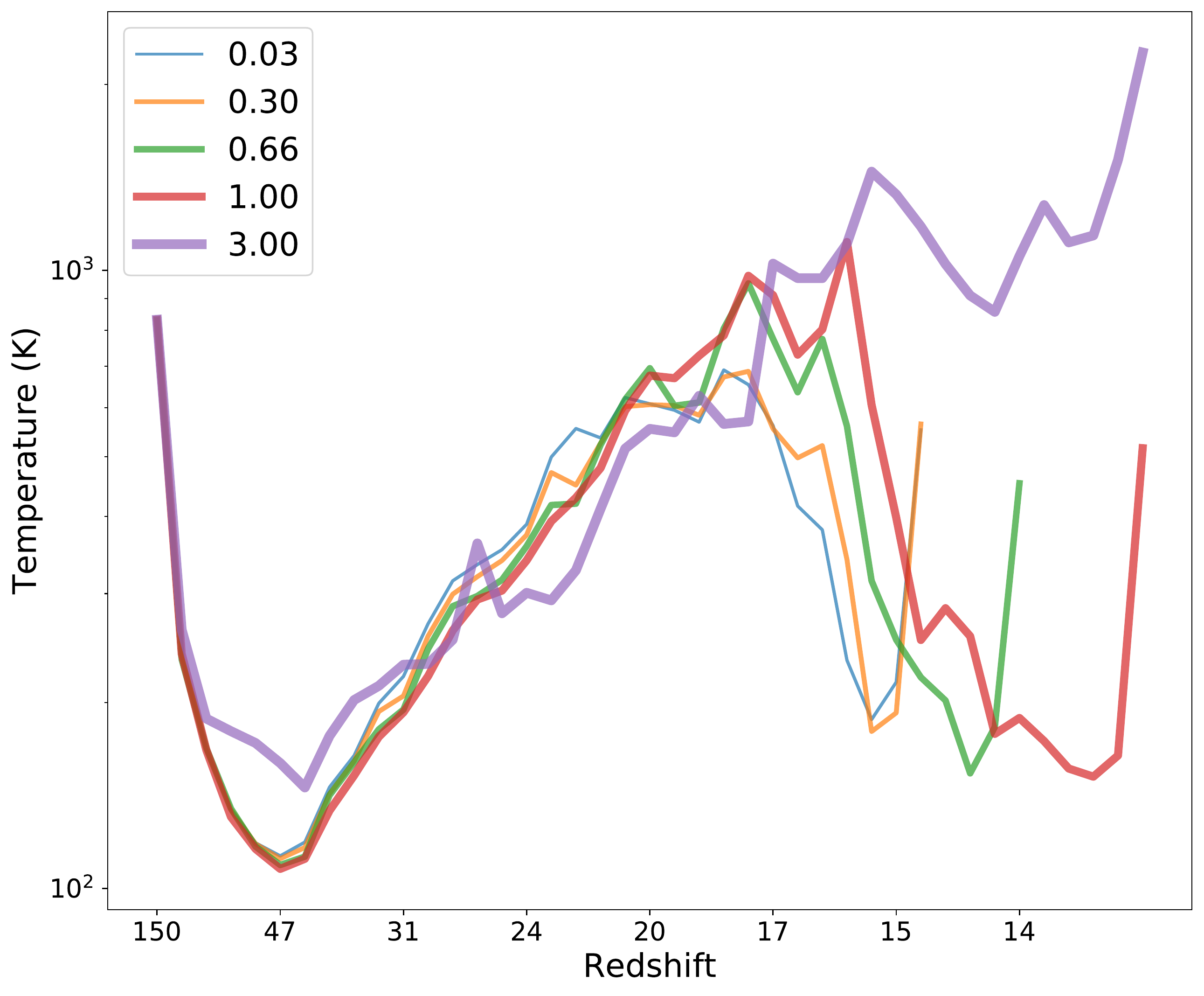}
	\caption{Plot of temperature at the cell with the highest density as a function of redshift for all simulations. In the super-sonic case, the central halo has yet to collapse and is continually heating up as it grows in mass. As the initial $v_a/c_s$ ratio increases, the peak temperature reached prior to cooling is higher, the time to collapse is delayed, and the duration of the cooling is extended.}
	\label{fig:maxT}
\end{figure}

Figure \ref{fig:maxT} shows the temperature evolution of the highest density point as a function of redshift for each of the different simulations. As the initial $v_a/c_s$ increases, the peak temperature reached by this point also increases. On the other hand, the minimum temperature reached by this densest point is lowered by a few 10s of Kelvin as the ratio is increased. This pattern does not hold true for the ratio of 3 as the object has yet to undergo collapse, but we would expect it to follow the pattern once it does collapse. This rise in temperature not only results in delayed collapse, but we also observe that, once collapse takes place, it is progressively elongated with increasing $v_a/c_s$ ratio.  
In the case of the super-sonic ratio of 3, the halo has not begun to cool even at the final data dump at z = 12.6. 

\begin{figure*}
	\includegraphics[width=\textwidth]{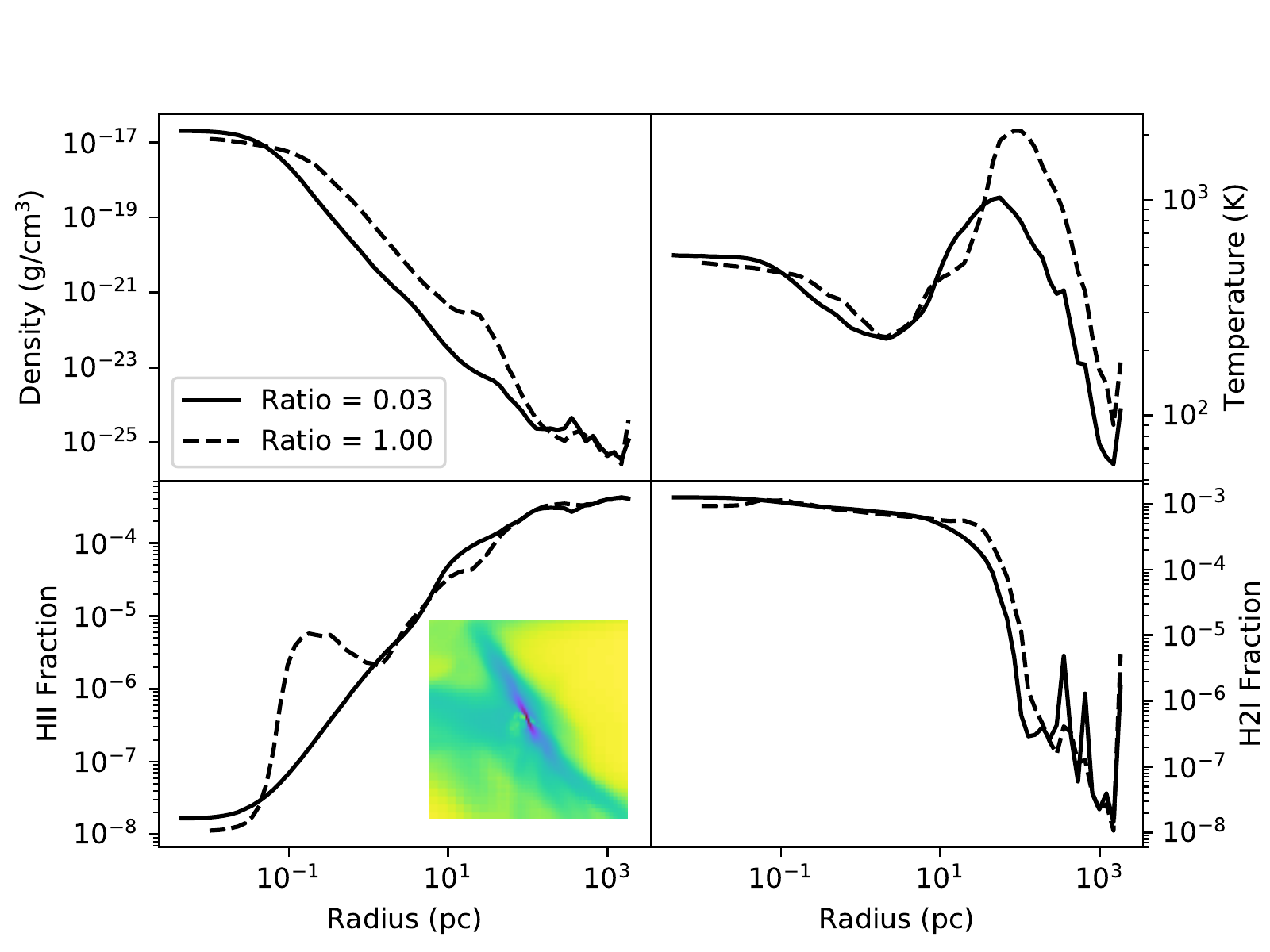}
	\caption{Radial profiles of density (upper left), temperature (upper right), HII fraction (lower left), and $\mathrm{H_2}$ fraction (lower right) centered about the densest point in the central halo for ratios of 0.03 (solid) and 1.00 (dashed). The trans-sonic case results in larger halo which produces the larger extended profile as seen in the density and temperature plots. Embedded in the HII fraction plot is a projection plot of the HII fraction weighted by density spanning 10pc centered around the densest point which highlights the highly asymmetric nature of the collapse.}
	\label{fig:prof}
\end{figure*}

Figure \ref{fig:prof} shows radial profiles centered around the central halo for a sub-sonic ratio of 0.03 and a trans-sonic ratio of 1.00. As the overall collapse is delayed, the halo is able to accrete more mass and thus we see a more extended profile in the density plot (upper left). The temperature profile (upper right) shows that the temperature in the core for the trans-sonic halo is a few 10s of K cooler but with a more extended heated tail. Surrounding the cool inner region is a hot gas which is heated to over 1000 degrees greater in the trans-sonic halo. As the profile is spherically-averaged, the actual temperatures in the heated region reach several thousand K. The neutral $H_2$ fraction (lower right) shows a corresponding extended profile for the trans-sonic case as the gas clouds that harbored the molecular cloud in the sub-sonic cases has been merged into the central object.

Inspecting the collapsed region in detail shows that the nature of the collapse itself has changed drastically between the two cases. In the sub-sonic cases, the central halo undergoes a mostly self-similar spherical collapse. On the other hand, the collapse in the trans-sonic halo proceeds along a particular axis along a filamentary structure. This results in a vastly different substructure particularly noticeable in the HII fraction profile. The ion fraction shows a pronounced increase by a couple orders of magnitude in the available ions in the cool region surrounding the central heated core. Figure \ref{fig:prof} also includes a projection of the ion fraction 
for the trans-sonic case as weighted by density spanning 10 pc to show the asymmetric nature of the collapse.

\begin{figure*}
	\includegraphics[width=\textwidth]{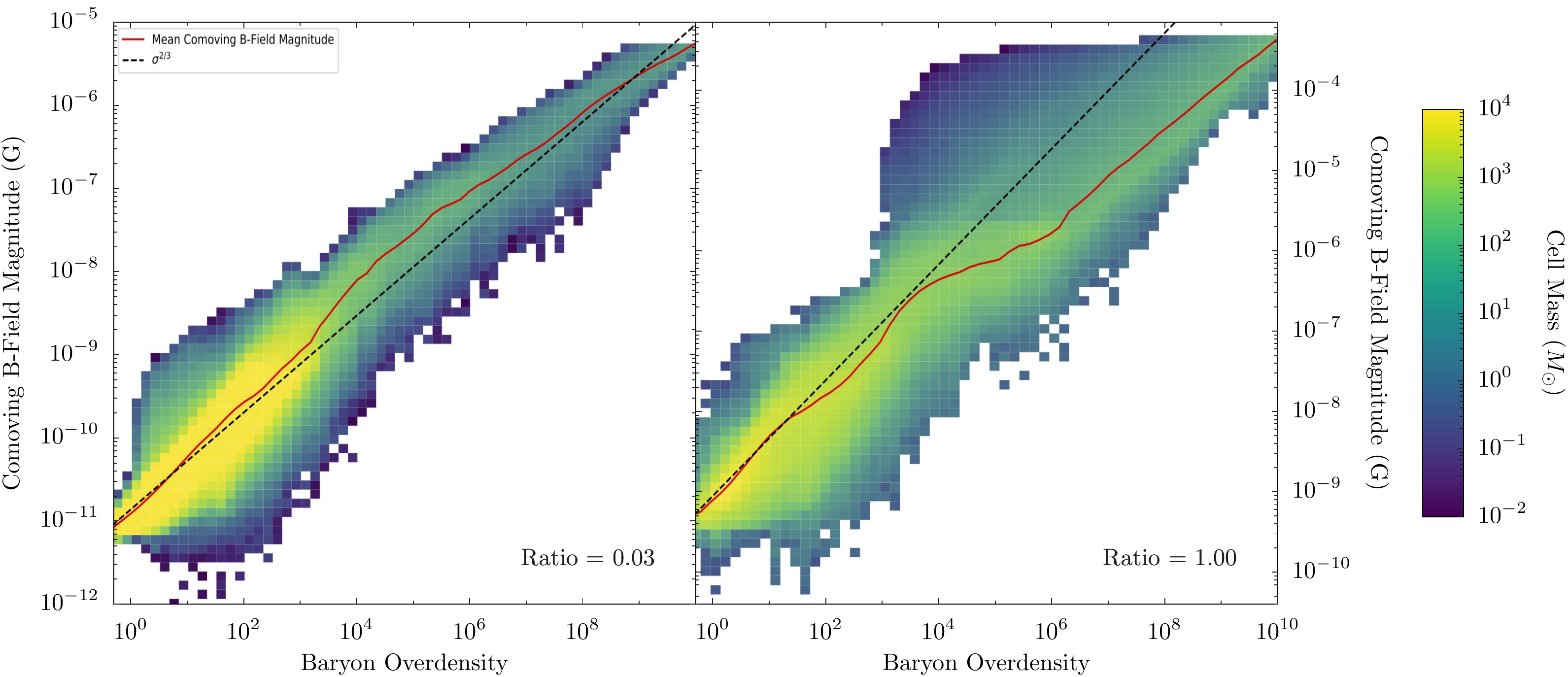}
	\caption{Phase plot of the comoving magnetic field strength against baryon overdensity within a sphere of 1 kpc surrounding the densest point. The left shows the sub-sonic ratio of 0.03 while the right shows 1.00. Red line indicates the average field strength while the dotted line is a $B \propto\rho^{2/3}$ reference trendline representing the ideal magnetized spherical collapse. While the sub-sonic case suggests some turbulent amplification, the trans-sonic halo is not even matching the compressional amplification. This further suggests that the gravitational collapse is occurring along the field lines.}
	\label{fig:rhoB}
\end{figure*}

\subsection{Magnetic Field Evolution}

We now present the behavior of the magnetic fields in our simulation. 

Figure \ref{fig:rhoB} shows a phase diagrams of comoving magnetic field strength vs the baryon overdensity in a sphere of radius 1 kpc surrounding densest cell in the trans-sonic run. The red line follows the mass-weighted average magnetic field strengths. The dotted line shows $B \propto\rho^{2/3}$, where $\rho$ is density, which is the scaling for the magnetic field amplification in the case of a ideal spherical collapse, or compressional amplification. In the sub-sonic scenario, there is still evidence of small scale dynamo driven amplification shown by the red line growing steeper than the dotted line, noting that it is muted relative to scenarios where $v_a <<$ 1.
On the other hand, in the trans-sonic case, the collapse is not driving strong amplification, and in fact the field is even less amplified than expected during ideal compression. The latter implies that much of the collapse must be occurring along the field lines, rather than squeezing lines together to drive amplification. This is supported by the projection plot in Fig. \ref{fig:prof}, where the collapse was no longer spherical and had a preferential axis. Furthermore, the field is effectively saturated having reached corresponding energies comparable to the kinetic energy in the system. And in fact, in this particular scenario, the magnetic energy is comparable to the kinetic component in this entire domain.

\begin{figure}
	\includegraphics[width=\columnwidth]{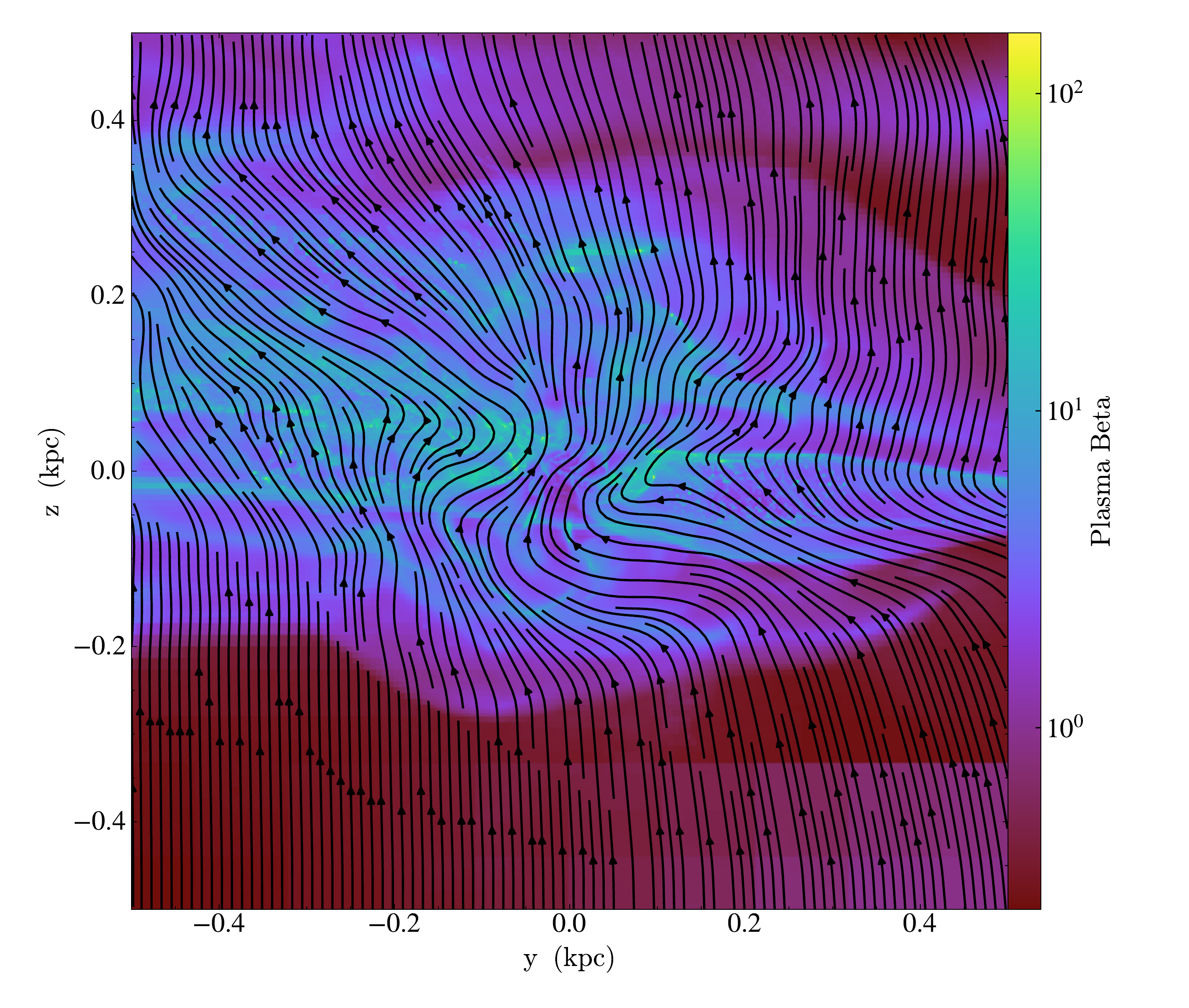}
	\caption{Projection plot of plasma $\beta$ weighted by density centered around the central object spanning 1 kpc across in the trans-sonic case. The streamlines trace the magnetic fields and shows the coherence of the initial magnetic field oriented in the z-direction over the span of the figure. In a large fraction of the total region, the magnetic pressure dominates over the thermal pressure.}
	\label{fig:beta}
\end{figure}

Figure \ref{fig:beta} shows a projection plot of the plasma $\beta$, which is the ratio of the thermal pressure to the magnetic pressure for the trans-sonic case. The magnetic fields are overplotted as streamlines and show that the initial magnetic field which was initialized in the z-direction is still coherent across the halo that is formed. Within the halo, the turbulent collapse of the gas pulls the field lines with it and reorders them.

\begin{figure}
	\includegraphics[width=\columnwidth]{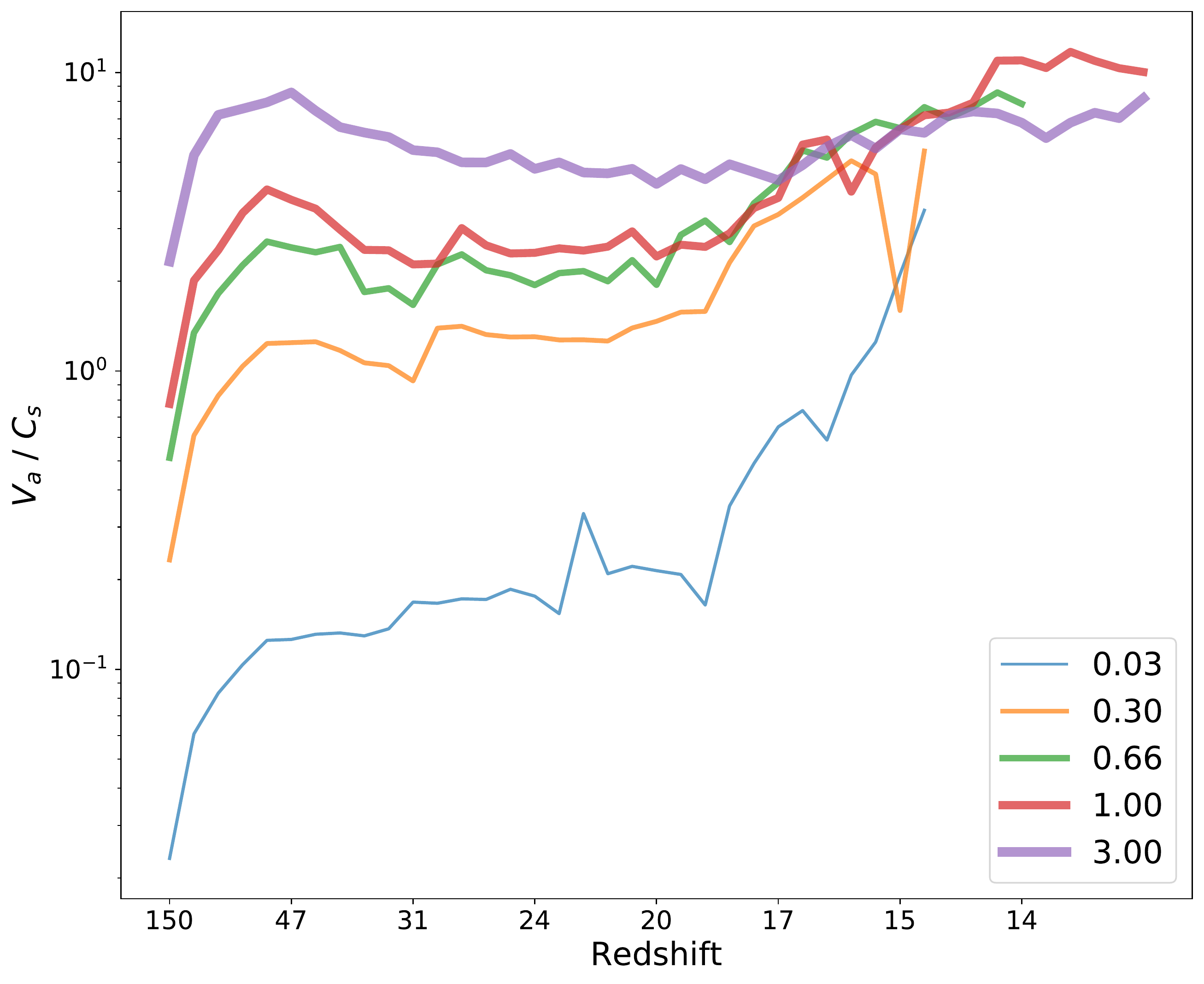}
	\caption{Plot of the $v_a/c_s$ ratio at the cell with the highest density as a function of redshift for all simulations. At sub-sonic ratios, the small-scale dynamo driven amplification rapidly increases this ratio. As the initial ratio is increased, the amplification is suppressed and the ratio near the center of the halo stays near $v_a/c_s \sim$10.}
	\label{fig:maxR}
\end{figure}

Figure \ref{fig:maxR} shows the $v_a/c_s$ ratio at the highest density point as a function of redshift for all the simulations. There is a ceiling for this ratio in the core of the central halo at $v_a/c_s \sim$10. For the sub-sonic cases, the magnetic field is rapidly amplified to approach this limit. As the initial field strength is increased, the degree of amplification is reduced overall and the ratio reaches a plateau. This corresponds again to the saturation point.

\section{Conclusions and Discussion} \label{sec:con}



In this {\em Letter}, we followed the collapse of the first mini-halos in cosmological simulations in the presence of primordial magnetic fields of comoving strengths in the range $\sim 2\times 10^{-12}-2\times 10^{-10}$G. We find that when fields are of order $5\times 10^{-11}$ G or larger, corresponding to the Alfv\'en speed being of the same order or larger than the sound speed at high redshift $z\simgt 100$, the formation of first stars in such mini-halos is significantly impacted.  In particular,


\begin{itemize}
    \setlength\itemsep{1em}
    \item With increasing initial magnetic field strength, the collapse of mini-halos due to the $H_2$-cooling is progressively delayed both in duration and final time of collapse. 
    \item Magnetic field amplification by the small-scale dynamo and by
simple flux-freezing during spherical collapse is increasingly reduced with
increasing field strength.    
    \item At high B-field strengths, $H_2$-cooling induced collapse can potentially be completely suppressed.
    \item Magnetic fields lead to asymmetric gravitational collapse
and an elevated ion population in the central few pcs of the mini-halo. 
    \item For all magnetic field strengths investigated,
the Alfv\'en- to sound -speed ratio in the center of the mini-halos
saturates at $v_a/c_s \sim$10, with 
saturation occurring earlier for stronger initial fields.
\end{itemize}

Our findings could have profound implications for early universe structure formation and beyond. First of all, the minimum collapse mass scale would be greatly sensitive to this effect as the magnetic fields suppress collapse in smaller mini-halos. This can impact the initial mass function of primordial stars and the resulting chemical evolution of the universe. The presence of larger pristine gas reservoirs by the time of collapse can result in radically different star formation scenarios. It can also perhaps more readily facilitate more exotic formation scenarios such as direct-collapse black holes.

We caution though, that our study has not included all relevant effects.
Apart from the neglect of ambipolar diffusion, a serious limitation is the
neglect of enhanced baryonic density perturbations induced by the magnetic
fields themselves. When taken into account, it may well be that, rather than
being delayed, star formation may occur earlier than in the no initial 
magnetic field counterpart. This is due to the enhanced power on small scales
making mini-halos collapse earlier. However, we believe that most other effects found in this study should remain and possibly lead to more massive first stars.

To place our findings into context, either inflationary produced PMFs or phase transition produced PMFs could relieve the
Hubble tension. In both cases a pre-recombination
field strength of $\sim 5\times 10^{-11}$G is required. However whereas in the former scenario this field strength is kept
to the present epoch, in the latter scenario fields are subject to further 
damping down to $1\times 10^{-11}$G through the epoch of recombination. 
Coincidentally this is approximately
the strength required to explain cluster magnetic fields to be entirely
primordial.
We may therefore conclude that only inflationary produced PMFs may influence
first structure formation, whereas phase transition produced fields are too
weak to have significant impact.




\acknowledgments

This work was performed using the open-source Enzo and YT codes, which are the products of collaborative efforts of many independent scientists from institutions around the world. Their commitment to open science has helped make this work possible.
This work was supported in part by the U.S. Department of Energy SLAC Contract No. DE-AC02-76SF00515.

%


\software{Enzo \citep{2014ApJS..211...19B},  
          YT \citep{2011ApJS..192....9T}}






\bibliography{paper.bib}{}
\bibliographystyle{aasjournal}



\end{document}